\documentstyle[12pt,epsf]{article}

\begin{document}

\newcommand{\nc}{\newcommand}
\nc{\beq}{\begin{equation}} \nc{\eeq}{\end{equation}}
\nc{\beqa}{\begin{eqnarray}} \nc{\eeqa}{\end{eqnarray}}
\nc{\lsim}{\begin{array}{c}\,\sim\vspace{-21pt}\\< \end{array}}
\nc{\gsim}{\begin{array}{c}\sim\vspace{-21pt}\\> \end{array}}
\nc{\el}{{\cal L}} \nc{\D}{{\cal D}}

\renewcommand{\thepage}{\arabic{page}}
\begin{titlepage}

\begin{center}

Final version \hfill YCTP-P7-00\\

 To appear in {\it Class. Quant. Grav.} \hfill hep-th/0006093\\

\bigskip

\bigskip

\vskip .5in

{\Large \bf Nonsingular deformations of singular
compactifications,}

\smallskip

\smallskip

{\Large\bf
 the cosmological constant, and the hierarchy problem}

\vskip .8 in

{\large Alan Chodos, Erich Poppitz, Dimitrios Tsimpis}

\vskip 0.2in

 {\tt alan.chodos, erich.poppitz, dimitrios.tsimpis@yale.edu}

\vskip 0.2in {\em
  Department of Physics,
  Yale University\\
  New Haven,
   CT 06520-8120, USA}

\end{center}

\vskip .4in

\begin{abstract}
We consider deformations of the singular ``global cosmic string"
compactifications, known to naturally generate exponentially large
scales. The deformations are obtained by allowing a constant
curvature metric on the brane and correspond to a choice of
integration constant. We show that there exists a unique value of
the integration constant that gives rise to a nonsingular solution.
The metric on the brane is  $dS_4$  with an exponentially small
value of expansion parameter. We derive an upper bound on the brane
cosmological constant. We find and investigate more general
singular solutions---``dilatonic global string"
compactifications---and show that they can have nonsingular
deformations. We give an embedding of these solutions in type IIB
supergravity. There is only one class of supersymmetry-preserving
singular dilatonic solutions. We show that they do not have
nonsingular deformations of the type considered here.
\end{abstract}
\end{titlepage}

\setcounter{page}{1}

\renewcommand{\thefootnote}{\#\arabic{footnote}}
\setcounter{footnote}{0}

\baselineskip=18pt
\section{Introduction and summary.}

The proposal that we live on a three-dimensional extended object
(``brane") embedded in some higher dimensional spacetime
\cite{ADD}, \cite{olderpapers}, \cite{RS} has recently attracted a
lot of attention. Such a scenario, for example with a pair of
millimeter-size extra dimensions, is not excluded by current
observations \cite{ADD}, provided the electroweak and strong
interactions do not feel the presence of the extra dimension.
Furthermore, the required localization of gauge interactions has a
very natural realization in string theory provided the extended
object is a $D$-brane \cite{JP}.

Gravity, on the other hand, is free to propagate in the extra
dimensions and becomes four dimensional only at distances large
compared to their size. The strength $M_{Pl}$ of four-dimensional
gravitational interactions is related to the strength $M$ of the
higher-dimensional gravitational theory: $M_{Pl}^2 \sim M^4 V_2$,
where $V_2$ is the volume of the transverse space and we assume
two large dimensions. Remarkably, if $M$ is in the (multi) $TeV$
range, and the size of the extra dimensions is a (sub) millimeter,
the above relation yields the correct value of $M_{Pl}\sim 10^{19}
GeV$. Thus,  in this ``large extra dimensions" scenario \cite{ADD}
the hierarchy problem appears as the question: ``Why are the extra
dimensions so large?"

It is  therefore interesting to look for ways to generate
exponentially large distances between objects of codimension two.
Potentials in two dimensions vary logarithmically and it is natural
to expect the appearance of exponentially large scales \cite{AB},
\cite{D}, \cite{AHall}.

The solutions of the coupled Einstein-Maxwell equations outside
charged point particles in two spatial dimensions \cite{Deser}
exhibit some relevant properties---particles of charge $g$ are
surrounded by a Killing horizon at an exponentially large distance
$\sim e^{{\cal{O}}(1){8 \pi^2\over g^2}}$ from the particle. In
the case of extended objects (``branes") in codimension two,
charged under appropriate rank antisymmetric tensor fields, there
is instead a mild\footnote{As $r$ approaches $r_*$ the curvature
scalar is finite, but the square of the Riemann tensor diverges,
see eqns.~(\ref{scalarcurvature}, \ref{riemannsquared}).} naked
singularity at an exponentially large distance, $M r_* \sim
e^{{\cal{O}}(1){8 \pi^2\over g^2}}$. This was first demonstrated
in \cite{CK1} in the context of global cosmic strings in four
dimensional spacetime and generalized in \cite{CK2} to a charged
extended object of codimension two in a spacetime of arbitrary
dimension.

The singular properties of charged 3-brane solutions of
codimension two have been exploited in \cite{CK2} to suggest an
interesting direction for solving the hierarchy problem: the
charged 3-brane creates a singularity at an exponentially large
distance, of order $e^{{\cal{O}}(1){8 \pi^2\over g^2}} M^{-1}$.
The assumption that spacetime terminates at the singularity leads
to a finite, exponentially large, four-dimensional $M_{Pl}$
without the usual fine tuning of parameters.

It may be that the singularity is indeed harmless and the
appropriate boundary conditions---no flow of conserved quantum
numbers through the singularity---are imposed by the fundamental
theory of quantum gravity.\footnote{In references
\cite{AHDKS},\cite{KSS} a similar situation appears, with a
(stronger) singularity at finite distance, for a codimension-one
brane in five dimensions. An analysis of the nature of such
singularities was made in \cite{G}.}
  However, one of the main points we wish to stress is that it is
important to study nonsingular deformations of the singular
solutions. Of particular interest are deformations that appear as a
choice of integration constants rather than parameters in the
lagrangian. It appears natural that the ``correct"---from the point
of view of the fundamental quantum gravity theory---choice of
integration constant would be the one that avoids the singularity;
the case is even stronger if there is a unique such value.

A class of deformations of the singular solutions of \cite{CK2},
obtained by varying the bulk cosmological constant $\Lambda_6$ was
 studied in \cite{RG1999}. It was shown that a unique nonsingular
solution exists if the bulk cosmological constant  is precisely
tuned to a negative (corresponding to an $AdS_6$ vacuum) value,
whose modulus is bounded above by an exponentially small scale:
$|\Lambda_6| < e^{- {\cal{O}}(1){8 \pi^2\over g^2}} M$. Thus, the
price to pay for obtaining a nonsingular solution in this case is
an enormous fine tuning of a parameter in the lagrangian.

In this paper, we study another class of deformations. We  allow a
constant curvature metric on the 3-brane---$dS_4$, $AdS_4$, or
$R_{1,3}$. This class of deformations appears as a choice of
integration constant  $\alpha$ \cite{RS}. We show, using the
methods of \cite{RG1996}, \cite{RG1995}, that there exists a {\it
unique} value of the brane cosmological constant $\alpha$ for which
the solution is nonsingular. The induced metric on the brane is de
Sitter with an exponentially small cosmological constant. The
singularity is replaced by a horizon at finite proper distance from
the brane.\footnote{The resulting spacetime is
 geodesically incomplete, as is the case in many recently considered
 ``compactifications" on noncompact spaces of finite volume.
 One has to either impose appropriate boundary conditions
 at the horizon, see \cite{BF}, \cite{RSalt}, \cite{GS},
 or complete the space by adding ``image" charges as in, e.g.
\cite{AHall}.}

We study deformations of two classes of singular charged
three-brane solutions. In Section \ref{ckdeformation} we study the
deformations of the ``global cosmic string" three-brane solution
of \cite{CK2}. In Section \ref{dilatondeformation} we find a more
general solution including also a dilaton field with an arbitrary
value of the dilaton coupling. This is motivated in part by
supersymmetry: for a particular value of the dilaton coupling the
dilatonic charged solutions include the singular BPS solutions
described in \cite{Duff}.

We investigate the possibility that the nonsingular solutions we
find solve both the hierarchy and cosmological constant problems;
further comments along these lines are given in the concluding
section and in ref.~\cite{Luty}. Our qualitative approach suffices
to show that there is a unique value of the brane cosmological
constant yielding a nonsingular solution. However, it allows us to
obtain only an upper bound on the value
 of $|\alpha|$. (To obtain a precise value one must proceed
numerically.) Even though exponentially small, this bound is still
many orders of magnitude larger than the experimental bound on the
cosmological constant; a  value of the bound closer to observation
can be obtained, but at the cost of fine tuning.

In Section~\ref{IIB}, we embed the solutions in type IIB
supergravity. We show that, within our embedding, the warped
solutions preserve no supersymmetry and that the only
supersymmetric solution is the unwarped solution of
ref.~\cite{Duff}. Using the analysis of
Section~\ref{dilatondeformation}, we show that this singular BPS
solution has no nonsingular deformation (supersymmetric or
otherwise) of the class considered here.

Finally, even though this paper is devoted to investigating the
``global cosmic  string" compactification, we note that similar
considerations---choosing integration constants to avoid
singularities---can be applied to the ``local cosmic string" case
considered in \cite{GS}. There the solution of the 6d Einstein
equations around an uncharged (in our terminology) three brane in
$AdS_6$ was found. The authors of \cite{GS} only considered the
case where the brane cosmological constant  $\alpha$  vanishes. In
fact, it can be shown, using the results of \cite{CP}, that the
choice of vanishing brane cosmological constant is the only one
leading to a nonsingular solution.  Outside   the core, the
solution of \cite{GS} is a particular case of the general
rotationally invariant metric around a three brane in $AdS_6$ found
in \cite{CP}: \beq \label{gsmetric} d s_6^2 = z^2 d s_4^2 +
f^{-1}(z) dz^2 +   f(z) d \theta^2~, \eeq where   $f \sim z^2 - a +
{ b \over z^3}$, where $a$ and $b$ are constants of integration,
with $a \sim \alpha$ (the brane cosmological constant, see
\cite{CP} for more details). Ref.~\cite{GS} considered the choice
of integration constants $\alpha = a = b = 0$, i.e. $f \sim z^2$.
The Minkowski three-brane is  the only choice where the metric
(\ref{gsmetric}) has no singularity at the horizon $z = 0$ (the
core of the solution is at $z = 1$, see eqn.~(34) of \cite{GS}).
This follows from the expression for the curvature,
\beq \label{gscurvature} R \sim f^{\prime \prime} - 8 {f^\prime
\over z} + 12 {f \over z^2}~, \eeq and from similar expressions for the
other curvature invariants.

\section{Action, ansatz, and equations of motion.}
\label{generalities}

 The action we consider is that of gravity in six dimensions,
 with the usual Einstein  action, minimally coupled to a scalar
 field $\phi$ and a four-form antisymmetric tensor field $C_{(4)}$:
\beq \label{gravityaction} S_{bulk} ~=~ M^4 ~ \int d^6 y \sqrt{-g}
\left( R - {1\over 2}
\partial_M \phi \partial^M \phi - {1\over 2\times 5!}~e^{-a
\phi}~F_{(5)}^2 ~
\right)~. \eeq
 The indices $M, N = 1,...,6$, while $\mu,\nu = 0,...,3$ and
$i,j = 1,2$ span the rest of the space; the metric has signature
$(-,+,\ldots, +)$ and $F_{(5)} = d C_{(4)}$ is the field strength
of the four form $C_{(4)}$. In what follows, we will consider
solutions both with ($a \ne 0$) and without ($\phi = a = 0$) scalar
fields.

We have in mind ``matter", in the form of $3$-branes, whose action
is proportional to the area of the world surface  they sweep in
the six-dimensional spacetime and include a topological coupling
to the four-form: \beqa \label{matteraction} S_{matter}&=& - f^4
\int d^6 y \int d^4 \sigma ~\delta^6(y - X (\sigma))~ \sqrt{-
\tilde{g} e^{2 a \phi}} \left( 1 + \ldots \right) \\
 &-& T^4 \int d^6 y \int d^4 \sigma ~\delta^6(y - X (\sigma))~
{1\over 4!}~
 C_{M_1...M_4}\epsilon^{\mu_1...\mu_4}
 X^{M_1}_{,\mu_1} ... X^{M_4}_{,\mu_4} ~, \nonumber
\eeqa where  $X^M (\sigma)$ is the embedding of the world surface
in space time (a sum over the various branes is needed if there is
more than one brane) $\tilde{g}_{\mu \nu} := X^M_{,\mu} X^N_{,\nu}
g_{MN}(y)$ is the induced metric and $\tilde{g}:= \det
\tilde{g}_{\mu \nu}$ (the terms with the dots denote other
possible matter constrained to the brane). $M$ and $f^4$ are the
six-dimensional Planck scale and brane tension, respectively. We
note that the fields $\phi$ and $C_{(4)}$ in (\ref{gravityaction})
are dimensionless; the charge of the $3$-brane under the
antisymmetric tensor field is then seen to be proportional to
  $(T/M)^4$.

A comment is due on the relevance of the ``matter" action
(\ref{matteraction}). One way to think about the brane action is as
describing the zero-thickness limit of a solitonic solution of a
six-dimensional field theory. Generally, however, in codimension
two and higher, the zero-thickness limit of defects coupled to
gravity---with the energy momentum tensor becoming a
distribution---is not well defined; see ref.~\cite{Geroch} for a
detailed discussion.\footnote{The action (\ref{matteraction}) will
be useful in one instance---the BPS dilatonic solution of
\cite{Duff} matches precisely to the distributional source
(\ref{matteraction}) for $a = 2$, $T = f$. The  equations of motion
of all bulk fields (metric, four-form, and dilaton) are satisfied
at all values of $r$, including the origin; in addition because of
the no-force condition, the brane equation following from varying
(\ref{matteraction}) with respect to the embedding coordinates is
also obeyed. We will discuss the supersymmetry properties of this
solution as well as its $\alpha \ne 0$ deformation in Section
\ref{IIB}.} We will think, instead, of the solutions we investigate
as describing the metric outside the core, of   size $r_0 \sim
M^{-1}$, of a solitonic solution; we will not deal in
 detail with matching to the source. The solution of \cite{CK1},
\cite{CK2} was found as the metric outside the core of such a
soliton; the solution where only the four form  is nonvanishing
($\phi = a = \alpha = 0$) is obtained after a  duality
transformation (in the bulk) of their solution.

We are looking for a solution which is a warped product of a four
dimensional space of constant curvature (Minkowski, de Sitter, or
Anti de Sitter) and has an $SO(2)$ isometry acting on the
transverse directions. The most general form of the metric,
consistent with these symmetries, is: \beq \label{metricansatz} d
s^2 ~=~ e^{2 A (t) } ~g^{(4)}_{\mu \nu}(x)  ~dx^\mu dx^\nu ~+
~e^{2 \beta(t) }~( d t^2 + d \theta^2)~, \eeq where $t = \log r$
and we use dimensionless units; all dimensions are restored at the
end by inserting appropriate powers of $M$.
 The functions $A$ and $\beta$ as well as the scalar field $\phi$ are
assumed to depend on the radial coordinate $t$  in the transverse
dimensions, while the four-dimensional metric $g^{(4)}_{\mu \nu}
(x)$ obeys: \beq \label{einsteineqns} R^{(4)}_{\mu \nu} - {1\over
2} g^{(4)}_{\mu \nu} R^{(4)} = \alpha g^{(4)}_{\mu \nu}. \eeq We
will not constrain the sign of the parameter $\alpha$ at this
point. Our conventions for the Einstein tensor are such that
$\alpha < 0$ ($\alpha > 0$) corresponds to (anti) de Sitter space.

The following ansatz, consistent with the symmetries of
(\ref{metricansatz}), for the non-vanishing components of the
five-form field strength satisfies both the equations of motion
and the Bianchi identity: \beq \label{fieldstrength} F^{\mu_1
\mu_2 \mu_3 \mu_4 t}~ =~ e^{a\phi(t)} ~ { \epsilon^{\mu_1 \mu_2
\mu_3 \mu_4 t \theta} \over \sqrt{-g} }\partial_{\theta}\chi;
\,\,\,\,\, \chi:=d_1 \theta, \eeq where $d_1$ is a constant of
integration, related to the charge of the ``brane" by Gauss's law.
The rest of the equations are:  \beqa \label{eqns} R^{(6)}_{\mu
\nu}-{1 \over 2}g^{(6)}_{\mu \nu}R^{(6)} &=&-{1\over
4}g^{(6)}_{\mu \nu} e^{-2\beta}(\phi'^2+8qe^{a\phi})  \nonumber ~,
\\
R^{(6)}_{tt}-{1 \over 2}g^{(6)}_{tt}R^{(6)} &=&{1\over
4}(\phi'^2-8qe^{a\phi})  \nonumber~,
\\
R^{(6)}_{\theta \theta}-{1 \over 2}g^{(6)}_{\theta \theta}R^{(6)}
&=&-{1\over 4}(\phi'^2-8qe^{a\phi})  ~,
\\
0&=& 4 A' \phi'+\phi'' -4qae^{a\phi} \nonumber~, \eeqa where the
prime refers to differentiation with respect to $t$ and $q :=
{d_1^2\over 8} > 0$ is related to the square of the charge of the
solution. The equations of motion in the form (\ref{eqns}) will be
useful in Section~\ref{IIB} when studying the embedding of our
solutions in type IIB supergravity.

In order to study the $\alpha \ne 0$ deformation of the singular
solutions, it is convenient to cast the equations of motion in
first order form and study the flow of the resulting dynamical
system. To this end, we introduce the ``time" $\tau$ \beq
\label{tau} \tau (t) ~=~ \int\limits^{t} d t~ {e^{\beta (t)} \over
e^{A (t)} }~. \eeq and the variables \beqa W ~&=&~ e^{A - \beta} A'
\equiv \dot{A}  \nonumber~,
\\
 V ~&=&~ e^{A - \beta} \beta' \equiv
\dot{ \beta}~, \\
  S ~&=&~ e^{A - \beta} \phi' \equiv \dot{\phi}   \nonumber~,
\eeqa where the dot refers to differentiation with respect to
$\tau$. In terms of the above variables the equations become first
order: \beqa \label{equationsofmotion} \dot{W} ~ &=& ~ - 3 ~W^2 - V
W - \alpha \nonumber ~,\\ \dot{V} ~ &=& ~ - V^2 + 12 ~W^2 + 5 ~V W
- {1 \over 2} S^2 + 4 \alpha ~,\\ \dot{S} ~ &=& ~- 3 ~S W - S V - 8
a~ W V - 12 a~ W^2 + {a \over 2}~ S^2 - 4 ~a~ \alpha \nonumber~,
\eeqa subject to the constraint, which is consistently propagated
in $\tau$ by the equations of motion (\ref{equationsofmotion}):
\beq \label{constraint} 3~ W^2 + 2 ~W V - {1 \over 8} ~S^2 +
\alpha ~=~ - q ~e^{2 A - 2 \beta + a \phi}~. \eeq

The equations of motion will allow us to study both the $\alpha =
0$ and $\alpha \ne 0$ solutions in terms of the flow of the
dynamical system of $W, V, S$ described by
(\ref{equationsofmotion}). The solutions of interest have, in
addition, to respect the constraint (\ref{constraint})---in
particular, only the trajectories of the $W, V, S$ dynamical
system for which the l.h.s. of (\ref{constraint}) is negative
solve the full set of the Einstein/scalar field equations.

To study the singularity structure of the solutions of
eqns.~(\ref{equationsofmotion}, \ref{constraint}), it is useful to
give some of the curvature invariants of the metric
(\ref{metricansatz}) on the solutions of the equations of motion.
For the scalar curvature, we obtain (for the de Sitter case,
$\alpha = - 3 H^2$; we   only need to consider this case, see
Section~\ref{ckdeformation}): \beq \label{scalarcurvature} R = 4
e^{- 2 A} \left( 3 H^2 + {1\over 4} S^2 - 2 V W - 3 W^2 \right) = 4
q e^{a \phi - 2 \beta} + {1\over 2} S^2 e^{- 2 A} ~, \eeq while,
for example, the square of the 6-d Riemann tensor is:
\beqa
\label{riemannsquared}
  R_{A B C D} R^{A B C D} &=&
% &~& {4 \over 3} e^{- 4 \beta} \left(
%62 q^2 e^{2 a \phi} - q e^{a \phi} ({7\over2} \phi'^2 - 128 A' \beta')
%+ 80 A'^2 \beta'^2 - 4 \phi'^2 A' \beta' + {7\over 33} \phi'^4
% \right)~ \\
 {248 \over 3} q^2 e^{2 a \phi - 4 \beta} - {4 \over 3} q e^{a
\phi - 2 A} \left({7\over2} S^2 - 128 V W\right) \nonumber \\
& & +{4 \over 3} e^{-4 A} \left( 80 V^2 W^2 - 4 S^2 V  W + {7\over
33} S^4
\right)~. \eeqa
It is useful to note that, similarly, all
curvature invariants can, on the solutions of the equations of
motion (\ref{equationsofmotion}, \ref{constraint}), be expressed
as functions of $V,W, S$. From (\ref{riemannsquared}) it follows
that a solution where these variables go to infinity for some
values of $\tau$ will be singular, provided $e^{- 2 A}$ does not
vanish fast enough.

More generally, we note that the phase space analysis of Sections
\ref{ckdeformation}, \ref{dilatondeformation} shows that either all
$W,~V,~S$ flow to infinity or all stay finite. Thus, trajectories
  for which $W,~V,~S$ all flow to infinity
 can be approximated, for $|W|, |V|, |S| \gg |\alpha|$,
 by the trajectories  of the system
(\ref{equationsofmotion}, \ref{constraint}) with $\alpha$ set to
 zero. As we wee will see shortly, in this case the equations can be
solved exactly and give rise to singular solutions. We therefore
conclude that for a nonsingular solution $W,~V,~S$ have to be
bounded.

\section{Singular nondilatonic charged solution: Cohen-Kaplan solution
 and its nonsingular deformation.} \label{ckdeformation}

The Cohen-Kaplan (CK) solution describes the spacetime outside the
core of a global ``cosmic string" and in our notations corresponds
to ignoring the dilaton and setting $\alpha=0$ in
(\ref{equationsofmotion}, \ref{constraint}). The equations can
then be integrated to give: \beq \label{cksolution} F(V,W):=~{1
\over W}~\left({ - W \over 2V +3W}\right)^{3\over 16}
\exp\left({-V \over 8W}\right)~=~ const. \eeq We can also solve
parametrically for $A$ and $\beta$ in terms of the variable $t$:
\beqa \label{parametricsolution} e^{A}&=&\left(1-{4 \over c} ~t
\right)^{1\over 4} \nonumber~,
\\
e^{\beta}&=&\left(1-{4 \over c} ~t\right)^{-{3\over 8}}~
\exp\left( -q \left( t^2 - {c \over 2} ~t \right) \right)~, \eeqa
where  $q = {d_1^2\over 8}$ (see eqn.~(\ref{fieldstrength})) and $c
> 0$ is a constant. By comparing (\ref{parametricsolution}),
(\ref{cksolution}), the value of the constant on the r.h.s. of
(\ref{cksolution}) is found to be: \beq \label{constant}
    - 3 q^{- {3 \over 16}} ~c^{5 \over 8} \exp \left( {q c^2 + 3 \over
16} \right)~.
\eeq
 The constants $q, \, c>0$ can be
fixed by  considering the core of the soliton to which
(\ref{parametricsolution}) matches, as explained in
Section~\ref{generalities}; the parameter $u_0$ of \cite{CK2} is
related to our parameters as $c^2 q = {8\over u_0}$. A singularity
occurs at an exponentially large distance $rM = e^{t} \sim
e^{c/4}$. (Recall that we take
 $M^{-1}$ as the characteristic size of the ``core'').

\begin{figure}

\caption{ \small The flow diagram of the equivalent dynamical
system (\ref{equationsofmotion}). The thin solid straight lines
represent the lines (\ref{separatricesa}) where the velocities
($\dot{W},
\dot{V}$) of the undeformed,  $H=0$, system  change sign, while the
two thicker (undashed) hyperbolae are the corresponding lines
(\ref{separatricesb}) of the deformed, $H\ne 0$, system. The signs
of  the velocities ($\dot{W}, \dot{V}$) in the various regions are
shown by $(\pm,\pm)$. The four critical points are where the two
hyperbolae intersect. A trajectory which satisfies the constraint
(\ref{constraint}) must lie between the two branches of the
invariant hyperboloid (represented by a thick dashed curve). The
invariant hyperboloid is also a critical trajectory, repulsive in
 the $(-H, 0)$ and attractive in the  $(H, 0)$ critical point,
 as indicated by the arrows.
 The singular CK solution starts in the  $(+, -)$ part of the $(W<0,
V>0)$ quadrant, turns around and goes to infinity in the $W,V$
plane asymptoting the
 $V = - {3\over 2} W$ line. Its nonsingular deformation is the
 attractive trajectory
 ending in the $(-H,0)$ critical point. The singular CK trajectory is
 always to the left of its nonsingular deformation. The singular
 unwarped ($W = 0$) BPS trajectory is given by the vertical axis. }

\vspace{1.2in}

{\centerline{ \epsfxsize=9in \epsfbox{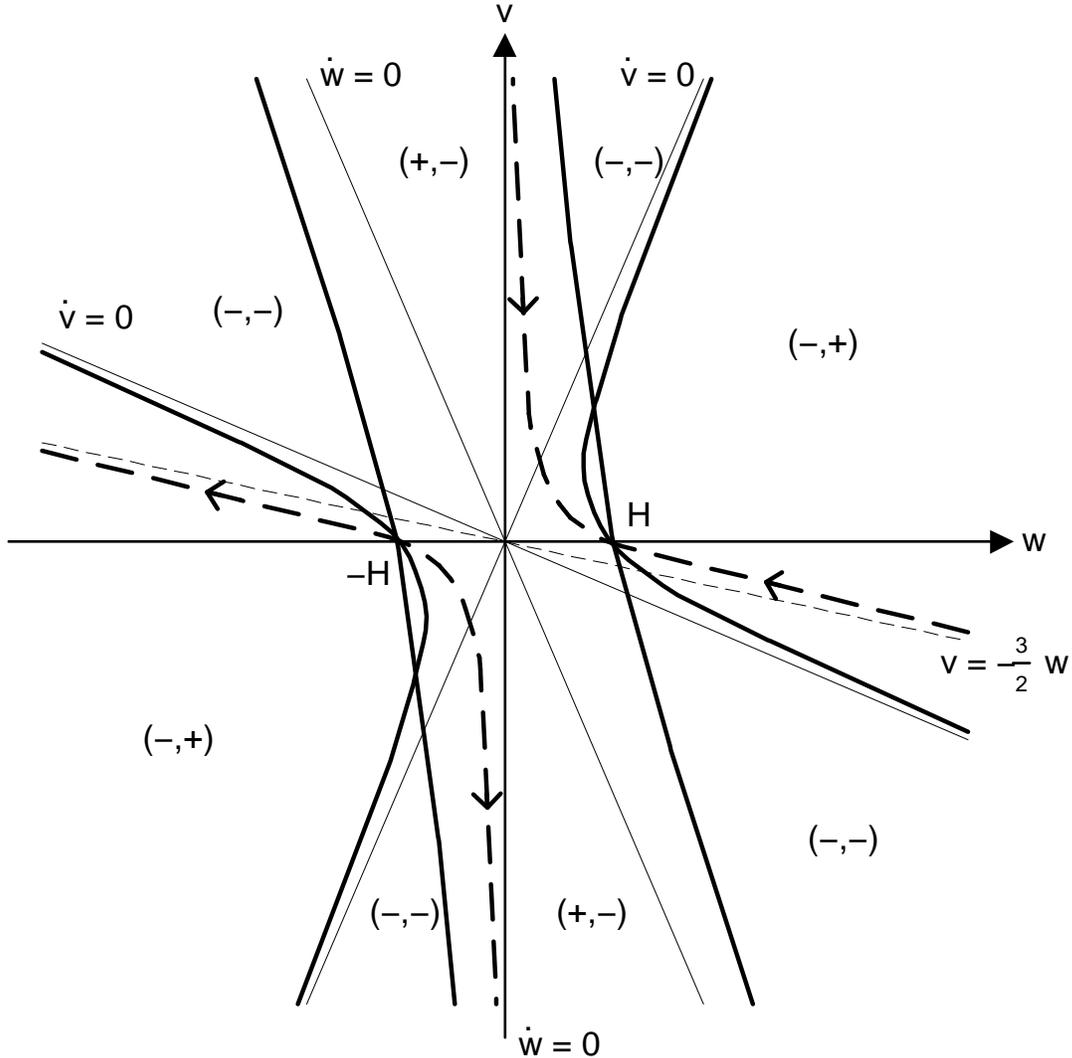} } }

\end{figure}

That a singularity appears is also clear from considering the
flow diagram (Fig. 1) of the first-order-form equations
 (\ref{equationsofmotion}), (\ref{constraint}) in the $W-V$ plane.
To draw it, note that the lines \beqa \label{separatricesa} W = 0
; \,\,\,\,\,\,\, V + 3 W &=& 0 \nonumber
\\
V=   {5 \pm \sqrt{73} \over 2 }  &W& \eeqa correspond to the lines
where velocities change sign, i.e. to $\dot{W}=0$, $\dot{V}=0$
respectively. The CK solution asymptotes the line $V+{3 \over
2}W=0$ as $t \rightarrow {c\over 4}$. Note also that the constraint
implies that the allowed trajectories lie in the region bounded by
the lines ${W=0, \,\, V+{3 \over 2}W=0 }$. In fact in view of the
positivity of $c, \, q$ the solution obeys $W<0, \, V>0$.

Now,  in view of the remarks following (\ref{riemannsquared}) any
solution that flows to infinity will be singular (note also that,
in the $W < 0, V>0$ quadrant $A$ is a decreasing function and the
factor $e^{- 2 A}$ is an increasing function of time). A
nonsingular solution, if it exists, should correspond to a
trajectory that flows to a critical point. First we note that the
system (\ref{equationsofmotion}) has no critical points for
$\alpha
> 0$ ($AdS_4$ case). In the $dS_4$ case ($\alpha<0$) there are four
critical points: $\{ W=V=\pm \sqrt{{|\alpha | \over 4}} \}$ and
$\{ V=0; \,\, W=\pm \sqrt{{|\alpha | \over 3}} \}$. Only the last
two obey the constraint.  Moreover, as is clear from investigating
the flows, only trajectories terminating at the $\{ W=-
\sqrt{{|\alpha | \over 3}}; \,\, V=0 \}$ critical point can be
nonsingular deformations of the CK solution and in the following
we will focus on this critical point.

For  $\alpha < 0$, however, it is not easy to integrate the
equations (\ref{equationsofmotion}, \ref{constraint}) any more.
Nonetheless we can still draw a flow diagram on the $W-V$ plane
(shown on Figure 1). Now $ \dot{W}=0$, $\dot{V}=0$ correspond to
the hyperboloids
\beqa
 \label{separatricesb}
 \left(W+{V \over
6}\right)^{2} - {V^{2} \over 36 }&=& {|\alpha| \over 3} \nonumber
\\
 {73 W^{2} \over 4} - \left( V- {5W \over 2} \right)^{2} &=&  4 |\alpha|
\eeqa
which at infinity asymptote the lines (\ref{separatricesa}).
The constraint now implies that the allowed trajectories lie in
the region between the two branches of the hyperboloid $(W+{V
\over 3})^{2}-{V^{2} \over 9 }= {|\alpha| \over 3}$.

We now define \beq \label{definitionofH} H:=\sqrt{{|\alpha|\over
3}}, \eeq where $H$ is the expansion parameter of the $dS_4$ metric
on the brane. A linear analysis near the $\{ W=- H; \,\, V=0 \}$
critical point reveals that it is a saddle. The attractive
trajectory, which corresponds to the nonsingular solution,
approaches the critical point with slope $dV/dW=-8$ (the repulsive
trajectory is, on the other hand, simply the invariant hyperboloid
of the dynamical system (\ref{equationsofmotion}) and is given by
the constraint equation (\ref{constraint}) with $q = 0$).

 The metric (and a ``Kruskal" extension) of the nonsingular solution
 can be obtained
explicitly in the limit $\tau \rightarrow \infty$, i.e. near the
critical point. In this way one establishes the existence of a
horizon, which replaces the singularity. To see this, note that
the metric near the critical point can be found by solving
$\dot{A} = - H$, $\dot{\beta} = 0$. Using a closed spatial section
parametrization of $dS_4$, we obtain: \beq \label{horizonsolution}
d s^2\big\vert_{crit} \simeq y^2  \left( - d \eta^2 + {\rm cosh}^2
\eta ~ d \Omega_3^2 \right) +
  d y^2 + e^{2\beta_0} d \theta^2~,
\eeq where $\beta \rightarrow \beta_0$ and $y := {1\over
H}e^{-H\tau} \rightarrow 0$ as the nonsingular trajectory
approaches the critical point $\tau \rightarrow \infty$. This
metric can be written as: \beq \label{horizonextended} d
s^2\big\vert_{crit} \simeq   - d T^2  + d R^2 + R^2 d \Omega_3^2 +
e^{2\beta_0}  d \theta^2 ~, \eeq by changing coordinates $T = - y
~{\rm sinh} \eta,\, R = y ~{\rm cosh} \eta$. Thus, the original
spacetime (\ref{horizonsolution}) corresponds to the region outside
the light cone of (\ref{horizonextended}), $R^2 > T^2$, and can be
extended through the horizon $y = 0$ to $y < 0$. The near-horizon
metric (\ref{horizonextended}) is approximately that of
five-dimensional Minkowski space times a circle of radius
$e^{\beta_0}$.\footnote{At the critical point (horizon), the
curvature invariants do not vanish, but are proportional to $q$, as
follows from eqns.~(\ref{scalarcurvature}), (\ref{riemannsquared}).
The $q$-dependence of the metric  only comes in  through higher
orders in the expansion around the critical point, which were
neglected in deriving (\ref{horizonsolution}).}

The precise matching to a core solution, as discussed in
Section~\ref{generalities}, is a difficult and model-dependent
problem. We will only make some qualitative remarks. The initial
conditions, given by the core solution, say at $t \sim 0$, are
given by a one-parameter curve (e.g., for $|W(0)|= c^{-1} \ll 1$,
by $W(0) V(0)
\simeq - {q\over 2}$) in the $(+,-)$ region of the
$(W<0,V>0)$ quadrant (see Fig.~1), sufficiently far from the
origin, so that the solution can still be
 approximated by the CK trajectory.
Now, as one varies $H$,
the attractive trajectory stays strictly to
the left of $H$ (since it lies entirely in the $(+,-)$ region,
where the $V$ velocity is
 positive) and
 therefore crosses the initial condition line
 always to the left of $H$. On the other hand,
 the attractive
  trajectory approaches the vertical
 axis in the $H=0$ limit;
 the only trajectory that goes through the critical point
 is then the  vertical axis. Thus, by continuity,
  one can always find
 a value for $H$ such that the nonsingular trajectory ending at
 the critical point passes through the given point on the
 initial condition line.

While obtaining the precise value  of $H$ that gives rise to a
nonsingular solution is only possible numerically, an upper bound
on $H$ can be obtained from the qualitative analysis of the flows.
From this analysis, it is clear that the CK solution is always to
the left of its nonsingular deformation. In particular, at the
point $\{ V_{0}, W_{0}\}$ where the CK trajectory intersects the
$\dot{V}=0$ line, we have \beq
\label{thebounda} |W_{0}| = {2  V_{0} \over
\sqrt{73}-5}
> H~.
\eeq Note that $H$ has dimensions of mass, since $\alpha$ has
dimensions $({\rm mass})^2$. Plugging the above into
(\ref{cksolution}), (\ref{constant}),  we obtain the inequality:
\beq \label{theboundb} q^{3\over 16} c^{-{5\over 8}} \exp\left(-{q
c^2 \over 16}\right) > {H \over M}~. \eeq To understand the
phenomenological significance of this bound, let us reduce the
Einstein term in the six-dimensional action \beq
\label{einsteinterm} M^4 \int{d^6x \sqrt{g_{6}}R_{6}} = M^4 I
\int{d^4x \sqrt{g_{4}}R_{4}} := M^2_{Pl} \int{d^4x
\sqrt{g_{4}}R_{4}}, \eeq where \beq \label{definitionofI} I:={1
\over M^2} \int{dt  ~e^{2 A + 2 \beta}} \eeq and $M_{Pl} \sim
10^{19}~ GeV$ (the long-distance-theory Planck scale), $M \sim 1
~TeV$ (the short-distance-theory Planck scale). Assuming that $I$
will not differ significantly from its CK value, we obtain: \beq
\label{CKvalueofI} I \sim {1 \over M^2}~ c^{1\over 4} q^{-{3\over
8}} \exp\left( {q c^2 \over 8}\right) = {M_{Pl}^2 \over M^4} \eeq

We will  only consider the case when the values of $c, q$ are not
fine tuned and the hierarchy of scales is explained by the
exponential factor in (\ref{CKvalueofI}) alone. Then
eqns.~(\ref{CKvalueofI}, \ref{einsteinterm}) imply: \beq
\label{theexponent} \exp\left({q c^2 \over 8}\right) \sim
\left({M_{Pl} \over M}\right)^2 \eeq
{}From equations (\ref{theexponent}, \ref{theboundb}) we then
obtain the following upper bound on the expansion parameter of the
$dS_4$ metric: \beq \label{theoreticalbound} H < \left({M \over
M_{Pl}}\right)^2 M_{Pl} \sim 10^{-32} M_{Pl} \eeq On the other
hand, the experimental bound on the vacuum energy density, $H^2
M_{Pl}^2 < \rho_{crit} \sim 10^{-120} M_{Pl}^4$, gives \beq
\label{experimentalbound} H  < 10^{-60} M_{Pl}~. \eeq In other
words, the experimental bound on $H$ (\ref{experimentalbound}) is
thirty orders of magnitude stricter than the upper bound we have
established in (\ref{theoreticalbound}). In terms of vacuum energy
density the
 upper bound (\ref{theoreticalbound}) is set by the
fundamental  gravity scale, $M$, and is of order $M^4$. We stress
that this is only an {\it upper} bound on the cosmological
constant (recall that we also used  the CK trajectory instead of
the nonsingular trajectory to obtain the value of $I$ for our
bound). To obtain the value of $H$ needed to obtain a nonsingular
solution, one would have to proceed with a numerical analysis of
the nonsingular deformation of the CK trajectory; we leave this
for future work.

\section{Singular charged dilatonic solution and
its nonsingular deformation.} \label{dilatondeformation}

Let us now include the dilaton in our discussion. We start with
$\alpha=0$ in which case the equations of motion
(\ref{equationsofmotion}) can be integrated. We demand that the
solution reduce to the CK one in the $a \rightarrow 0$ limit. The
result is: \beqa \label{generalizedCK} A&=&{1 \over 4} \log\left(1
- {4 \over c} ~t\right) \nonumber
\\
\beta&=&{q c^2 \over 16} + {3 \over 8} \log {c \over 4}  +{1
\over4}\left(u_0+{8-d \over a^2}\right)\log\left({c \over 4}
-t\right) + {2 \over a^2} \log\left(1-{q a^2 \over 2}\left({c
\over 4} -t\right)^{d\over4}\right)
\\
\phi&=&\phi_0 + a u_0 A - a \beta \nonumber, \eeqa where \beq d ~
:=~8~ \sqrt{\left({u_0 \over 4}+{3 \over 8}\right)a^2 + 1}~, \eeq
  $u_0 \geq -{3 \over 2}$ is an integration constant, and the constant
part of the dilaton, $\phi_0$,
   is fixed in terms of $c, q, u_0$ by   the constraint equation
(\ref{constraint}). The range of $t$ is \beq \label{rangefort} {c
\over 4} \geq t \geq {c \over 4} - \left({2 \over q
a^2}\right)^{4\over d} \eeq It is straightforward to check that
indeed in the $a \rightarrow 0$ limit (\ref{generalizedCK})
reduces to (\ref{parametricsolution}).

The solution (\ref{generalizedCK}) is a ``dilatonic''
generalization of the singular CK solution and is itself singular.
Curvature invariants blow up at both limits of the range of the
variable $t$; as in Sect.~\ref{ckdeformation} we imagine that near
the lower limit of the range (\ref{rangefort}) of $t$ our solution
is matched to a smoooth solitonic solution to which
eqn.~(\ref{generalizedCK}) is a long-distance approximation.

As in the previous section, we cannot integrate analytically the
equations of motion for $\alpha \neq 0$. However,  one can still
analyze the three-dimensional flow diagram in the space $W,V,S$.
Proceeding as before one establishes the existence of a unique
nonsingular solution ending at the critical point $\{ V=S=0; \,\,
W=-\sqrt{{|\alpha| \over 3}}\}$. The critical point corresponds to
an approximately $R^{1,4} \times S^1$ horizon, exactly as in the
non-dilatonic case.

A bound on $H$ can be established as before. The integral $I$,
defined in (\ref{definitionofI}), can here also be calculated in a
closed form, with the assumption that its value for $\alpha \neq
0$ is not significantly different from its value for $\alpha=0$.
The result of integrating over the whole range (\ref{rangefort})
is: \beq \label{dilatonicI} I= {1\over d} ~\left({c\over
4}\right)^{{1\over 4}} \exp\left( {q c^2 \over 8} \right) \left(
{2 \over q a^2}\right)^{c_0} ~ B \left(c_0,1+{4\over a^2}\right)
\eeq where \beq c_0:={2\over d}\left({8-d\over a^2}+u_0+3\right)~.
\eeq It is easy to check that the above $I$ reduces to the value
for the CK solution (\ref{CKvalueofI}) in the $a \rightarrow 0$
limit. The phenomenological analysis of the previous section goes
through virtually unchanged. For example, eqn.~(\ref{theboundb})
gets replaced, upon repeating the analysis of the flows for the
$V, W, S$ dynamical system, by: \beq \label{dilatonicH} {H \over
M} < c^{- {5 \over 8}} \exp\left(-{ q c^2 \over 16}\right) x_*^{ -
{c_0 \over 2}  } \left( 1 - {q a^2 \over 2} x_* \right)^{- {2
\over a^2}}~, \eeq where $x_* =({ c\over 4} - t_*)^{d\over 4}$,
with $t_*$ given by solving the equation $\dot{V} = 0$. Although
rather complicated looking  once the explicit form of  the
solution (\ref{generalizedCK}) is plugged into $\dot{V} = 0$ (see
eq.~(\ref{equationsofmotion})), this equation determines $x_*$ as
an algebraic function of the parameters $q, a, u_0$. Exponentially
large or small numbers can appear only upon fine tuning these
parameters. Thus, comparing (\ref{dilatonicI}), (\ref{dilatonicH})
with (\ref{theoreticalbound}), (\ref{theexponent}), we see
that---with exponential accuracy---the bound on $H$ can not be
improved without fine tuning. (The limits of $a \ll 1$ and $a \gg
1$ are easiest to study analytically and support this conclusion.)

In other words, the inclusion of the dilaton cannot help obtaining
a stricter upper bound on the cosmological constant without fine
tuning. As in Sect.~\ref{ckdeformation}, obtaining the actual
value of $H$ for the nonsingular spacetime instead of an upper
bound is only possible numerically.

\section{Embedding in IIB supergravity.}
\label{IIB}

Here we will embed in type $IIB$ supergravity the solutions found
in the previous two sections. We show that the embedding does not
preserve any supersymmetry. Therefore, if the ``global cosmic
string" compactifications producing an exponential hierarchy can be
fully embedded in string theory (that is, if a string-derived model
for the core can be found---we only consider the embedding of the
exterior of the string in $IIB$ supergravity) they would correspond
to nonsupersymmetric string vacua.

Our strategy is first to consider a background for $IIB$ involving
the fields $\chi,~\phi,~g^{(6)}_{mn}$ of our six-dimensional
solution. We then show that on this background the $IIB$ equations
reduce to the equations of motion (\ref{eqns}). In this Section, we
use $M, N$ for the ten-dimensional indices and $\mu, \nu$---for the
4d indices. The fields of $IIB$ supergravity are the graviton
$g^{(10)}_{MN}$, the axion $\rho$ and the dilaton $\varphi$,
parametrizing an $SL(2, R)/U(1)$ coset space, a pair of two-forms
$B^{1,2}_{mn}$ which form an $SL(2, R)$ doublet, a four-form with
self-dual field strength $F_{(5)}$ and two complex Weyl fermions: a
gravitino and a dilatino.

We are interested in a type $IIB$ background  obeying the
six-dimensional equations of motion (\ref{eqns}). We will take the
fields to be independent of the last four coordinates ($I, J =
6,...,9$ in what follows). In addition, the ansatz should have the
same isometries as our six dimensional ansatz,
(\ref{metricansatz})---the $SO(1,3)\times SO(2)\times SO(4)$
isometry (for the warped Minkowski case), with the $SO(2)$ acting
as a shift of  $\theta$ and the $SO(4)$---the isometry of the four
extra coordinates.

We will consider an ansatz where only the metric, dilaton, and
axion are nonvanishing (one could instead consider a nonvanishing
four-form field; this is related to the case of nonvanishing axion
by T-duality in the $6...9$ directions). In a background where all
forms and all fermions are set to zero the equations of motion of
the type IIB supergravity fields are (these are the equations given
in \cite{GSW}, after $U(1)$ gauge fixing): \beqa \label{twobeqs}
R^{(10)}_{MN} - {1 \over 2 } g_{MN} R^{(10)}&=& {1\over 2}
\left(\partial_M \varphi \partial_N \varphi
-{1\over 2} g_{MN} \partial_K \varphi \partial^K \varphi \right)
+{1\over 2}e^{2\varphi} \left( \partial_M \rho \partial_N \rho
-{1\over 2}g_{MN} \partial_K \rho \partial^K \rho \right)
 \nonumber
\\
0 &=& \partial_M \left(\sqrt{- g_{(10)}}\partial^M\varphi \right)
- \sqrt{- g_{(10)}} e^{2\varphi}\partial_M\rho \partial^M\rho
\\
0 &=& \partial_M \left(\sqrt{- g_{(10)}}\partial^M \rho
\right) +2 \sqrt{- g_{(10)}} \partial_M \varphi \partial^M\rho
\nonumber~, \eeqa where $g_{(10)} = {\rm det} g_{(10)}$.
The most general ansatz for the nonvanishing components of the
ten-dimensional metric consistent with the above symmetries is:
\beqa
\label{gravitonansatz} g^{(10)}_{MN}&=& g^{(6)}_{MN};
\,\,\,\, M,N=0, \dots 5~,
\nonumber
\\
g^{(10)}_{IJ}&=& e^{\Psi(t)} \delta_{IJ}; \,\,\,\, I,J=6, \dots 9~,
\eeqa where $g^{(6)}_{MN}$ is as in equation (\ref{metricansatz}).
It is straightforward to show that the equations of motion
(\ref{twobeqs}) reduce to equations (\ref{eqns}) for \beq
\label{embeddings} a= 2;~ ~\Psi=0; ~~\varphi = \phi;~ ~\rho = d_1
\theta~~. \eeq

It is natural to ask whether the above embedding of our solution
preserves any supersymmetry in the case of warped Minkowski space
$g^{(4)}_{\mu\nu}=\eta_{\mu\nu}$. The supersymmetry transformations
of the dilatino and the gravitino are parametrized by a complex
Weyl spinor $\varepsilon$. The conditions for unbroken
supersymmetry in the background (\ref{embeddings})  with all fields
but the graviton and the scalars  set to zero are:
 \beqa
\label{susytransformations} &\Gamma^M \left( \partial_M \varphi+ i
e^{\varphi} \partial_M \rho
\right)~ \varepsilon^* ~=~ 0 \nonumber~,
\\
&\left( \partial_M + {1 \over 4} \omega_{MNL} \Gamma^{NL} +
{i\over 4} e^{\varphi} \partial_M
 \rho \right)~ \varepsilon~=~0~,
\eeqa where $\Gamma_{MN} ={1\over 2}  [\Gamma_M, \Gamma_N]$ and
$\omega_{MNL}$ is the spin connection: $$ \omega_{MNL} =  {1 \over
2} \left( e_{MA} ( \partial_N e^{A}_L - \partial_L e^A_N )
 - e_{LA} (\partial_M e^{A}_N - \partial_N e^A_M)
 - e_{NA}(\partial_L e^{A}_M - \partial_M e^A_L)\right)~.
$$ The only nonvanishing components of the spin connection are:
\beqa \label{spinconnection}
    \omega_{\mu \nu t} &=&  - \omega_{\mu t \nu} ~= - \eta_{\mu \nu}
e^{2 A} A^\prime , \nonumber
  \\
 \omega_{\theta t \theta} &=& - \omega_{\theta \theta t} ~=  ~e^{2
\beta} \beta^\prime .
\eeqa

Vanishing of the dilatino supersymmetry variation requires that the
parameter of the unbroken supersymmetry transformation
$\varepsilon$ obey (the gamma matrices in (\ref{susy1}),
(\ref{susy2}) have curved indices, i.e. are vielbein-dependent):
\beq \label{susy1} \left(    \phi^\prime ~ \Gamma_\theta \Gamma^t
+ i e^\phi  d_1\right) \varepsilon^* =0~, \eeq
while the gravitino variation  implies the conditions:
\beqa \label{susy2}
\partial_I ~ \varepsilon &=& 0~; ~~~I = 6  \ldots 9~, \nonumber \\
\left( \nabla_\theta - {1 \over 2} B^\prime~ \Gamma_\theta^{~~t} +
{i \over 4} e^\phi d_1 \right) \varepsilon &=& 0~,
\\
\partial_t ~ \varepsilon &=& 0~, \nonumber
\\ \left(
\partial_\mu - {1\over 2} A^\prime ~\Gamma_{\mu}^{~~ t} \right)
\varepsilon &=& 0 ~\nonumber~, \eeqa
where $\nabla_\theta \varepsilon :=\left( \partial_\theta - {1
\over 2}~\Gamma_\theta^{~~t} \right) \varepsilon$ is the
$\theta$-component of the covariant derivative in flat space and
$B$ is related to $\beta$ from (\ref{metricansatz}) by $B = \beta -
t$; the transverse part of the metric in $(r = e^t, \theta)$
coordinates is $e^{2 B}(dr^2 + r^2 d
\theta^2)$.
Along all  other directions the flat-space covariant derivative
 reduces to a simple derivative.
 The integrability condition following
 from the fourth equation implies that the factor $A^\prime = 0$
  if any supersymmetry is to be preserved in the Minkowski background.
We conclude that our dilatonic solution of Section
\ref{dilatondeformation}, which has a nontrivial warp factor $A$,
see eqn.~(\ref{generalizedCK}), does not preserve any
supersymmetry.

However, there exists another class of solutions, already known in
the literature \cite{Duff}. These unwarped ($A = const.$) singular
solutions  were shown in \cite{Duff} to obey a ``no-force"
condition---a probe brane with the same coupling to the dilaton and
antisymmetric tensor field as in (\ref{matteraction}) with $f = T$
 feels
no force towards the brane at $r = 0$. We will now obtain the
explicit form of these unwarped solutions by requiring unbroken
supersymmetry. To begin with, note that eqn.~(\ref{susy1}) imposes
a strong constraint on the form of the supersymmetry-preserving
solution: a $t$-independent $\varepsilon$ can obey (\ref{susy1}) if
and only if: \beq
\label{phicondition}
\phi^\prime ~=~ k d_1 e^\phi~,~ ~ k^2 = 1~. \eeq The last
condition on $k$ follows from rewriting equation (\ref{susy1}) as:
\beq \label{susy12}
\left( 1 - i k ~ \Gamma_\theta \Gamma^t \right)
\varepsilon^* = 0~, \eeq
and requiring the existence of a nontrivial solution. Note that
(\ref{susy12}) (for $k=\pm1$) is just the chirality projection in
the $t, \theta$ directions and half of the $\varepsilon$ components
are projected out. Now let us demand that $\varepsilon$ be a
covariantly constant spinor of flat ten dimensional Minkowski
space. Explicitly,
\beq
\label{covconst}
\partial_{\mu}~\varepsilon = \partial_I~\varepsilon
= \partial_t~\varepsilon = \nabla_\theta~\varepsilon~ = ~0~. \eeq
All supersymmetry equations are automatically obeyed except for the
second of eqns.~(\ref{susy2}) which requires, in view of
(\ref{phicondition}), that
\beq \label{susy22} \left(    2 i k B^\prime ~ \Gamma_\theta
\Gamma^t +  \phi^\prime \right) \varepsilon  = 0~. \eeq
Agreement with (\ref{susy12}) requires (taking a real
 $i  \Gamma_\theta
\Gamma^t$):
\beq \label{betacondition} B^\prime = - {1 \over 2} \phi^\prime ~,
\eeq
giving, together with (\ref{phicondition}) the solution of
\cite{Duff}, for $k = + 1$. In view of the comments following
(\ref{susy12}) it preserves half the supersymmetry of type $IIB$
supergravity, i.e. has sixteen real supercharges. In our notation,
as is easy to see, $W = 0$, $V = - {1 \over 2} S$ is a solution of
(\ref{equationsofmotion}), obeying the constraint
(\ref{constraint})  with $\alpha = 0$.

The above supersymmetric (unwarped and singular) solution is
represented (projected on the $W,V$-plane) on the flow diagram of
Fig.~1  by the vertical axis, with the direction of the flow
towards  negative infinity. The solution does not have a
nonsingular deformation with $\alpha \ne 0$. To see this, recall
first, as discussed in Section \ref{dilatondeformation}, that the
$W,V,S$ system
  has the same two critical points of interest as the $S
= a = 0$ system, and that these are the $(\pm H, 0)$ points
shown on Fig.~1. The directions of the flow velocities in Fig.~1
indicate that the singular BPS solution can not flow to the ($-H,
0$) critical point and could only be deformed to end at the ($H,
0$) critical point. However, it is easy to see that the attractive
trajectories at this critical point must lie on the invariant
hyperboloid (now a two-dimensional surface in the $W,V,S$ space),
as indicated on Fig.~1. The invariant hyperboloid obeys the
constraint equation with vanishing charge for all values of $t$ and
hence a trajectory that lies on it can not be the deformation of
the $q \ne 0$ nonsingular solution. We conclude that the $H
\ne 0$ deformations of the BPS trajectory continue to flow to
infinity and the solution remains singular.

We note that while this work was being completed, the paper
\cite{BHM} appeared, which discusses the embedding of dilatonic
``global cosmic strings" in string theory.

\section{Concluding remarks.}

We have demonstrated, both for the dilatonic and nondilatonic
singular ``global cosmic string" compactifications, the existence
of a unique value of the $dS_4$ expansion parameter $H$, for which
the singularity is absent and have given a strict upper bound on
the critical value of $H$. (One's hope is that the quantum gravity
theory chooses the value of the integration constant for which the
solution is nonsingular.) We have also shown that the field
configuration outside the core of the string can be embedded in
type $IIB$ supergravity and that it breaks all supersymmetry.

With exponential accuracy (without fine tuning parameters) the
upper bound---derived using a qualitative analysis of the
flows---on the corresponding vacuum energy density is of order
$M^4$, with $M$--the fundamental gravity scale ($TeV$, say).
Whether the actual critical value of $H$ can naturally be small
enough to agree with observations can only be decided numerically.
We should note, however, that even if the classical critical value
of $H$ is small enough, quantum loops (e.g., ``standard model"
quantum corrections to the ``brane tension") will probably change
its value, as one expects the critical value of $H$ to depend on
the ``brane" parameters. It is difficult, even adopting a specific
model for the core and ignoring bulk loop corrections, to estimate
the magnitude of this change, as an analytic expression for the
dependence of the relevant quantities---the critical value of the
de Sitter expansion parameter $H$ and the volume of transverse
space---on the brane tension is lacking.

\section{Acknowledgments.}

We want to thank G. Moore for useful discussions and suggestions.
We are also grateful to N. Arkani-Hamed, S. Ramgoolam, and M.
Shaposhnikov for discussions, and to G. Dvali and M. Luty for
comments on the manuscript.

\end{document}